\documentclass[twocolumn,secnumarabic, amsmath,amssymb, aps, pre,superscriptaddress]{revtex4-2}
\usepackage{graphicx}
\usepackage{bm}
\usepackage{multirow}
\usepackage{color}
\newcommand{\m}[0]{\rm{m}}

\newcommand{\mum}[0]{\mu\m}

\begin{document}
\title{Cellular-scale mechanism of cell crawling responding to substrate stiffness }%
\author{Sohei Nakamura}%
\email{nakamura.sohei@cmt.phys.kyushu-u.ac.jp}
\affiliation{Department of Physics, Kyushu University, Fukuoka 819-0395, Japan}
\author{Mitsusuke Tarama}%
\email{tarama.mitsusuke@phys.kyushu-u.ac.jp}
\affiliation{Department of Physics, Kyushu University, Fukuoka 819-0395, Japan}
\date{\today}%
\begin{abstract}
Biological cells are able to adapt their behaviour in response to environmental cues. 
Durotaxis is a phenomenon in which cells adjust their migration depending on the mechanical properties of a surrounding substrate. 
Although durotaxis has been studied more than two decades, basic cellular-scale mechanism of how cells regulate the motility responding to substrate stiffness remains to be elucidated. 
We address this issue by developing a theory utilising a mechanochemical model that integrates intracellular biochemical reactions with cellular deformation and substrate adhesion. 
Numerical analysis reveals that the characteristic speed and diffusion constant of cells change non-monotonically with respect to substrate stiffness, indicating the emergence of an optimal stiffness for migration. 
In addition, by introducing a memory effect that allows feedback from cell mechanics to the intracellular chemical reactions, the persistence time increases with substrate stiffness on a substrate softer than the optimal. 
We further investigate theoretically the origin of the non-monotonic dependence, that is comparable to the experimental observations, in terms of cell deformation and symmetry breaking in substrate adhesion. 
We believe that our study provides a unifying framework to understand complex durotactic cell migration. 
\end{abstract}
\maketitle
\section{Introduction} \label{sec:introduction}
The ability of sensing and responding to environments is of fundamental importance for animate systems. 
This is a function not only of higher multicellular organisms, but even a single cell can exhibit such an ability~\cite{Nakagaki2000Maze}. 
Examples include chemotaxis~\cite{king2009}, phototaxis~\cite{hegemann1991}, thermotaxis~\cite{poff1977}, gravitaxis~\cite{hader2017}, and Magnetaxis~\cite{delong1993}. 
While many theoretical studies focus on swimming cells, understanding the response dynamics of crawling cells is also important, in particular in the connection to that of multicellular organisms. 

Crawling cells are distinguished from swimming ones regarding the migration mechanism. 
While swimming cells stir the surrounding fluid to achieve locomotion, crawling cells adhere to the substrate or extracellular matrix and produce traction force to move forwards. 
Since cells migrate spontaneously, its locomotion is characterized by the force-free condition, which is known as the Purcell’s scallop theorem~\cite{purcell1977} for microswimmers at low Reynolds number. 
It is also true for crawling cells. 
In fact, the simple sum of the traction force is negligibly small~\cite{Tanimoto2014}, which results not only from the force-free nature but also from the fact that the inertia is negligibly small due to its typical size and velocity~\cite{Danuser2013Mathematical}. 
The basic mechanism of cell crawling under the force-free condition has been studied from various viewpoints~\cite{Leoni2017Model,Tarama2018}.

Among the response to environmental cues, durotaxis is a phenomenon that crawling cells exhibit in response to the substrate stiffness~\cite{lo2000}. 
Depending on the conditions, cells migrate towards the stiffer substrate region (positive durotaxis)~\cite{isenberg2009,tse2011,evans2018,duchez2019,kang2024} or towards the softer region (negative durotaxis)~\cite{isomursu2022,huang2022,singh2014} under the existence of the gradient in the substrate stiffness. 
Crawling cells can also respond to the pattern of different substrate stiffness~\cite{ebata2020,ebata2022} and to the dynamic change in the substrate stiffness~\cite{iwadate2013}. 
Durotaxis is found in various biological processes including development~\cite{shellard2021}, homeostasis~\cite{Dingal2015Fractal}, and disease~\cite{Goldmann2024Durotaxis}. 
Moreover, the substrate stiffness also affects differentiation~\cite{engler2006}. 

There are many studies modelling durotaxis, which can be separated into three groups~\cite{Shellard2021Durotaxis}: molecular-clutch model, adhesion-based model, and persistent-migration model. 
The molecular-clutch model considers the force-dependent characteristics of adhesion molecules, which strengthen the linkage between the cell and the substrate in response to mechanical tension on a stiffer substrate \cite{bangasser2013}.
The adhesion-based model takes into account the substrate adhesion on a more coarsened level corresponding to a focal adhesion or a group of them that matures and strengthen the substrate adhesion if the substrate is stiffer~\cite{Pallares2023}.
While the former two models includes either molecular or subcellular details to reproduce the cellular-scale behaviour, the persistent-migration model is a cellular-scale model standing on the assumption that cell migration persistence depends on the substrate stiffness~\cite{Novikova2017Persistence-Driven}. 
Although the connection between the molecular-clutch model and the adhesion-based model is rather straightforward, little is understood how the subcellular-scale mechanism of substrate rigidity sensing is related to the cellular-scale dynamics assumed in the persistent-migration model. 

In this paper, we address this issue by developing a theory based on a subcellular-element model for cell crawling and investigate the cellular-scale mechanism how cells can adapt migration depending on substrate rigidity. 
The subcellular-element model was developed by Newman to model the rheological property of biological cells~\cite{Newman2007}, and was later extended to cell crawling that is coupled to intracellular chemical reactions~\cite{Tarama2022}. 
In this paper, we further extend the mechano-chemical subcellular-element model to incorporate the influence of substrate stiffness by means of the adhesion-based method, as described in the next section. 
In section~\ref{sec:Results}, we analyse the dynamics that the model cell exhibits numerically and show the existence of optimal substrate stiffness for cell crawling. 
Then we investigate theoretically cellular-scale mechanism behind the non-monotonic dependence of cell crawling on substrate stiffness by focusing on the shape deformation and the symmetry breaking in substrate adhesion. 
Section~\ref{sec:Discussion} is devoted to discussion. 

\section{Mechano-chemical model}\label{sec:model}
We start by defining the mechano-chemical model for crawling cells that includes the influence of elastic substrate. 
Our model is twofold, namely mechanical and chemical parts, that are coupled through active force generation and substrate adhesion. 
We explain them one by one, and then introduce the impact of substrate stiffness. 
\begin{figure}[t]
\includegraphics[width=0.45\textwidth]{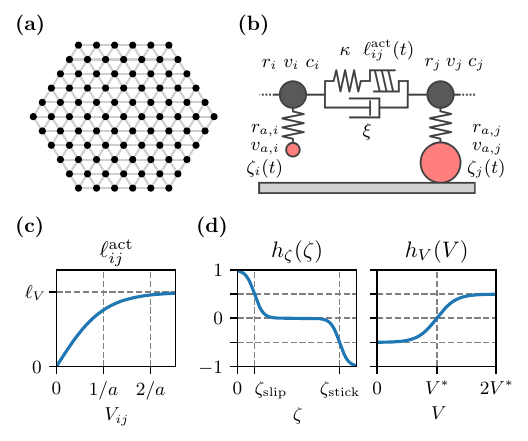}
\caption{Schematics of the mechanochemical model. (a) A single cell is described by a set of sebcellular elements (black) that are connected by visco-elastic springs (grey bars). (b) Visco-elastic spring connecting subcellular elements is of Kelvin-Voigt type, and includes a linear actuator that changes the length $\ell^{\rm act}_{ij}$ in time. Each subcellular element is connected by an elastic spring to an adhesion element (red), which changes substrate friction coefficient $\zeta_i(t)$ as schematically indicated by its size. (c) Dependence of $\ell^{\rm act}_{ij}$ on the chemical concentration $V_{ij}$. (d) Function form of $h_\zeta(\zeta)$ and $h_V(V)$ that determine the substrate friction coefficient $\zeta_i(t)$.}
\label{fig:fig1}
\end{figure}

The mechanical part of the model is constructed based on subcellular element method, where a single cell is modelled by several particles (i.e., subcellular elements) connected by visco-elastic springs of Kelvin-Voigt type that include linear actuators (Fig.~\ref{fig:fig1}a). 
In addition, we introduce adhesion elements, that represent cell-substrate adhesion (Fig.~\ref{fig:fig1}b). 
Adhesion element $i$ is connected to subcellular element $i$ by an elastic spring with elastic constant $k_a$ and zero rest length. 
Note that in the limit $k_a \to \infty$ the original model introduced in Ref.~\cite{Tarama2022} is obtained. 
The equation of motion for subcellular element $i$ is given by
\begin{align}
&\zeta^\prime \bm{v}_i + \sum_{j \in \Omega_i} \xi \ell_0 (\bm{v}_{i} - \bm{v}_{j}) \notag\\
&= \sum_{j \in \Omega_i} \frac{\kappa}{\ell_0} \hat{\bm{r}}_{ij} \left\{ r_{ij} - (\ell_0 + \ell_{ij}^{\textrm{act}}(t)) \right\} -k_a(\bm{r}_i - \bm{r}_{a,i}) +\bm{f}_i^{\textrm{area}},
\label{eq:Cell mechanics1}
\end{align}
where $\bm{r}_i$ and $\bm{v}_i$ are the position and velocity of subcellular element $i$, and $\bm{r}_{ij}  = \bm{r}_j -\bm{r}_i$. 
Here we have used the abbreviations $x = | \bm{x} |$ and $\hat{\bm{x}} = \bm{x} / x$. 
The summation is calculated over the set of subcellular elements $\Omega_i$ that are connected to subcellular element $i$. 
$\xi$ and $\kappa$ are the the intracellular damping rate and elastic modulus representing cellular viscosity and elasticity, respectively, and $\zeta^\prime$ is a (constant) substrate friction coefficient. 
$\ell^\textrm{act}_{ij}(t)$ denotes the length of the linear actuator that stretches and shrinks, leading to the active force $\bm{f}^{\rm{act}}_{ij} = -\kappa \hat{\bm{r}}_{ij} \ell^{\rm{act}}(t) / \ell_0$. 
The term including spring constant $k_a$ comes from the elastic spring that connects the subcellular element $i$ and the adhesion element $i$. 
The counterpart of this force enters in the equation of motion for adhesion element $i$ at position $\bm{r}_{a,i}$ with velocity $\bm{v}_{a,i}$, that is given by
\begin{align}
\zeta _{i}(t) \bm{v}_{a,i} &= k_a( \bm{r}_i -\bm{r}_{a,i})
\label{eq:Cell mechanics2}
\end{align}
with $\zeta_i(t)$ the substrate friction coefficient that changes its characteristics in time. 
The last term in eq.~\eqref{eq:Cell mechanics1} prevents the collapse of the subcellular element network on the course of stretching and shrinking of the visco-elastic spring, and is given by $\bm{f}_i^{\textrm{area}} = -{\partial U^{\textrm{area}}}/{\partial \bm{r}_i}$ with $U^{\textrm{area}} = \sum_{\left\langle j,k,\ell \right\rangle } \sigma_S /S_{jkl}^2$ and $\sigma_S = 10^{-6}$. 
Here, $S_{jkl} = (\bm{r}_{jk} \times \bm{r}_{jl}) \cdot \hat{\bm{e}}_z/2$ with $\hat{\bm{e}}_z$ the unit vector perpendicular to the substrate is the area of triangle $\langle j,k,\ell \rangle$ formed by the connected subcellular elements.

Note that in eqs.~\eqref{eq:Cell mechanics1} and \eqref{eq:Cell mechanics2} the inertia terms are omitted because of the typical size and speed of a cell that are of the order of $10\,\mum$ and $10\, \mum/\rm{min}$~\cite{Danuser2013Mathematical}. 
We also note that this model satisfies the force free condition, that is, the simple sum of the active force that the cell produces vanishes: 
\begin{align}
\sum_{i} \sum_{j \in \Omega_i} \bm{f}^{\rm{act}}_{ij} = \sum_{i} \sum_{j \in \Omega_i} \frac{-\kappa}{\ell_0} \hat{\bm{r}}_{ij} \ell^{\rm{act}}(t) = 0
\label{eq:force_free}
\end{align}
Together with the equations of motion \eqref{eq:Cell mechanics1} and \eqref{eq:Cell mechanics2}, this leads to vanishing simple sum of traction force: 
\begin{align}
\sum_i \left[ \zeta^\prime \bm{v}_i +\zeta _{i}(t) \bm{v}_{a,i} \right]= 0
\label{eq:force_free_substrate}
\end{align}

The chemical model is given by a Gray-Scott type reaction-diffusion equations that read
\begin{align}
\frac{\partial U_i}{\partial t} =& D_U \nabla^2 U_i + G_U(U_i,V_i)
\label{eq:Intracellular chemical reaction1}\\
\frac{\partial V_i}{\partial t} =& D_V \nabla^2 V_i + G_V(U_i,V_i)
\label{eq:Intracellular chemical reaction2}
\end{align}
where the reaction terms are given by
\begin{align}
G_U(U,V) =& -\frac{\alpha UV^2}{K_K + \left\langle V^2 \right\rangle} + \frac{\beta UV}{K_P + \left\langle U \right\rangle} -\gamma U + S
\label{eq:Intracellular chemical reaction3}\\
G_V(U,V) =& +\frac{\alpha UV^2}{K_K + \left\langle V^2 \right\rangle} - \frac{\beta UV}{K_P + \left\langle U \right\rangle} -\mu V
\label{eq:Intracellular chemical reaction4}
\end{align}
with the global couplings
\begin{align}
\left\langle U \right\rangle = \frac{1}{N} \sum_{i \in \Omega} U_i,~
\left\langle V^2 \right\rangle = \frac{1}{N} \sum_{i \in \Omega} V_i^2
\label{eq:Intracellular chemical reaction5}
\end{align}
The subscript $i$ of $U_i$ and $V_i$ represents the subcellular element index, and $\Omega$ and $N$ in eq.~\eqref{eq:Intracellular chemical reaction5} are the set of all subcellular elements and its total number, respectively. 
This set of reaction-diffusion equations is introduced to model the phosphorylate and de-phosphorylate reaction between phosphatidylinositol (4,5)-bisphosphate (PIP2) and phosphatidylinositol (3,4,5)-trisphosphate (PIP3) that is found to form travelling waves in \textit{Dictyostelium} cells, in which the concentration of PIP2 and PIP3 are given by $U$ and $V$, respectively~\cite{Taniguchi2013}. 
Note that in our model the concentrations $U$ and $V$ are existing only on the subcellular elements. 
To calculate the Laplacian in eqs.~\eqref{eq:Intracellular chemical reaction1} and \eqref{eq:Intracellular chemical reaction2}, we employ the moving particle semi-implicit method~\cite{Koshizuka1998};
\begin{align}
\nabla^2c_{i} = \frac{4}{\lambda}\sum_{j\neq i} \frac{1}{n_{ij}}\left(c_{j} - c_{i}\right) w(r_{ij}) 
\label{eq:Numerical analysis3}
\end{align}
where $c_{i}=\{U_{i},V_{i}\}$,  $\lambda=r_e^2/6$, and $n_{ij}=(n_{i}+n_{j})/2$ with $n_{i}=\sum_{i\neq j}w(r_{ij})$.
In this definition, $n_{i}$ corresponds to the number of neighbouring subcellular elements around the $i$th subcellular element that are weighted by
\begin{align}
w(r)=\left\{\begin{array}{ll}(r_e / r) - 1 & \textrm{if~} r < r_e \\0 & \textrm{otherwise}\end{array}\right.
\label{eq:Numerical analysis4}
\end{align}
with the cutoff length $r_e=4 \ell_0$.

The mechanical and chemical models are coupled through two intracellular activities: the actuator length $\ell^{\rm act}_{ij}(t)$ and the substrate adhesion characteristics $\zeta_i(t)$. 
Since PIP3 promotes the actin polymerization that leads to membrane protrusion, we assume 
\begin{align}
\ell^{\rm{act}}_{ij}(t) = \ell_0 \tanh{\left[ \pi V_{ij}(t) \right]} 
\label{eq:Cell elongation and contraction1}
\end{align}
so that $\ell_{ij}^{\rm act}$ increases with $V_{ij} = (V_i + V_j)/2$.
See Fig.~\ref{fig:fig1}c.
Although the cell-substrate adhesion is known to be controlled by the intracellular chemical substances~\cite{Schwarz2013}, its reaction pathway is not well understood yet. 
So, here we assume that depending on $V_i$ the substrate friction characteristic transitions between the two states, namely the adhered stick state with the coefficient $\zeta_{\rm stick}$ and the de-adhered slip state with $\zeta_{\rm slip}$;
\begin{align}
\tau_{\zeta}\frac{d\zeta_i}{dt} = h_{\zeta}(\zeta_i) - h_V(V_i)
\label{eq:Substrate stiffness1}
\end{align}
where $\tau_{\zeta}$ is the characteristic time scale and
\begin{align}
&h_{\zeta}(\zeta) = -\frac{1}{2}\tanh{\left[\frac{\zeta - \zeta_{\rm{stick}}}{\epsilon_{\zeta}}\right]} - \frac{1}{2}\tanh{\left[\frac{\zeta - \zeta_{\rm{slip}}}{\epsilon_{\zeta}} \right]}
\label{eq:Substrate stiffness2}\\&
h_V(V) = \frac{1}{2}\tanh{[\sigma_V(V-V^*)]}
\label{eq:Substrate stiffness3}
\end{align}
$\epsilon_{\zeta}$ represents the transition sharpness and $V^*$ is the threshold concentration (Fig.~\ref{fig:fig1}d).
Here, we set the parameters as $\tau_{\zeta_{\tau}} = 0.01$, $\epsilon_{\zeta} = \zeta_{\rm{slip}}/2$, $\sigma_{V} = 2\pi$, and $V^{*}=0.5$.

In addition, we introduce the polarity $\bm{P}(t)$ that memorizes the migration direction with degradation time $\tau_m$:
\begin{align}
\bm{P}(t) = \int_{-\infty}^{t} \bm{V}_{\rm{com}}(t') \frac{1}{\tau_{m}} \exp\left(-\frac{t-t'}{\tau_m}\right) dt' \label{eq:polarity2}
\end{align}
where $\bm{V}_{\rm{com}}$ is the centre-of-mass velocity
\begin{align}
\bm{V}_{\rm{com}} = \frac{1}{2N}\sum_{i}(\bm{v}_i + \bm{v}_{a,i}).
\end{align}
Depending on the polarity, chemical stimuli $(\delta U_i,\delta V_i) = (-I_{\rm excite},I_{\rm excite})$ are introduced periodically with the time interval $\delta t$ on a single subcellular element $i$ selected each time randomly with the probability given by
\begin{align}
p_{i}(t) = {\mathcal C}^{-1} \exp{\left[ C_m \left( \bm{r}_i - \bm{X}_{\rm{com}} \right) \cdot\bm{P} \right]} \label{eq:polarity4}
\end{align}
with the sensitivity $C_m$ and the normalization coefficient ${\mathcal C} = \sum_{i}\exp{\left[ C_m \left( \bm{r}_i - \bm{X}_{\rm{com}} \right) \cdot\bm{P} \right]}$ and the centre-of-mass position 
\begin{align}
\bm{X}_{\rm{com}} = \frac{1}{2N}\sum_{i}(\bm{r}_i + \bm{r}_{a,i})
\end{align}
This memory effect gives feedback from the cellular mechanics to the intracellular chemical reaction. 

Finally, we introduce the impact of substrate stiffness to the model. 
The molecules responsible for the substrate adhesion organise molecular complex called focal adhesion, which matures stronger on a stiffer substrate~\cite{espina2022}.
As a result, the cell-substrate adhesion gets stronger as the substrate stiffness increases~\cite{Moriyama2018}. 
Instead of considering the molecular details, for simplicity we phenomenologically include the influence of the substrate stiffness into the model through the strength of adhered stick state $\gamma_{\rm stick}$.
Namely, $\gamma_{\rm stick}$ takes a larger value for a stiffer substrate. 
Hereafter we quantify this by substrate stiffness parameter $\gamma_s = \zeta_{\rm{stick}} / \zeta_{\rm{slip}}$ ($> 1$), which increases with the substrate stiffness. 

To solve numerically the time-evolution equations~\eqref{eq:Cell mechanics1}, \eqref{eq:Cell mechanics2}, \eqref{eq:Intracellular chemical reaction1}, \eqref{eq:Intracellular chemical reaction2}, and \eqref{eq:Substrate stiffness1}, we employ the forth-order Runge-Kutta method with time increment $dt=2.5 \times 10^{-4}$. 
Following Ref.~\cite{Tarama2022},
the parameters are set as summarized in table~\ref{tab:Parameters}. 
The magnitude of the stimulus is set as $I_{\rm excite} = 1.0$.
Note that, although the stimuli is provided at the interval of $\delta t = 0.15$, not all the stimuli evolves into a travelling wave. 
As a result, the model cell migrates according to the intracellular chemical reaction wave, which deforms the cell shape and changes the substrate friction, as depicted in Fig.~\ref{fig:fig2}a. 
\begin{table*}[t]
\caption{
Values of (a)~mechanical and (b)~chemical parameters.
}\label{tab:Parameters}
\centering
\resizebox{\textwidth}{!}{
\renewcommand{\arraystretch}{1.3}
\begin{tabular}{|ll|l|l|l|l|l|}
\multicolumn{4}{l}{(a)\ Mechanical parameters} & \multicolumn{1}{l}{} & \multicolumn{2}{l}{(b)\ Chemical parameters} \\
\cline{1-4} \cline{6-7}
Model parameters & & Simulation values & Typical value & & Model parameters & Simulation values \\
\cline{1-4} \cline{6-7}
Number of cell element and adhesion element & $N$ & 91 & & &   $D_{U},D_{V}$ & 0.48\\
Bond length at rest  & $\ell_{0}$ & 0.1048 & $\sim 1\rm{\mu m}$ & & $\alpha$  & 240\\
Intracellular damping rate & $\xi$ & 0.004166 &$\sim 0.0004\ \rm{nN\cdot min(\mu m)^{-1}}$ & &   $\beta$ & 90\\
Intracellular elastic modulus & $\kappa$ & 10.00 & $100\rm{nN}$ & & $K_{K}$ & 5\\
Cell-adhesion bond spring constant & $k_a$ & 95.42 & $\sim95\ \rm{nN (\mu m)^{-1}}$ & & $K_{P}$ & 5\\
Substrate friction coefficient & & & & & $S$ & 30\\
\ of cell elements & $\zeta^\prime$ & 0.1099 & $\sim 0.1\ \rm{nN\cdot min(\mu m)^{-1}}$ & & $\gamma$ & 6\\
\ of adhesion elements in the slip state & $\zeta_{\textrm{slip}}$ & 0.1099 & $\sim 0.1\ \rm{nN\cdot min(\mu m)^{-1}}$ & & $\mu$ & 30\\
\cline{1-4} \cline{6-7}
\end{tabular}
\renewcommand{\arraystretch}{1}
}
\end{table*}

\section{Results}\label{sec:Results}
\subsection{Optimal substrate stiffness for cell migration}\label{subsec:result1}
\begin{figure*}[t]
\centering
\includegraphics[width=0.9\textwidth]{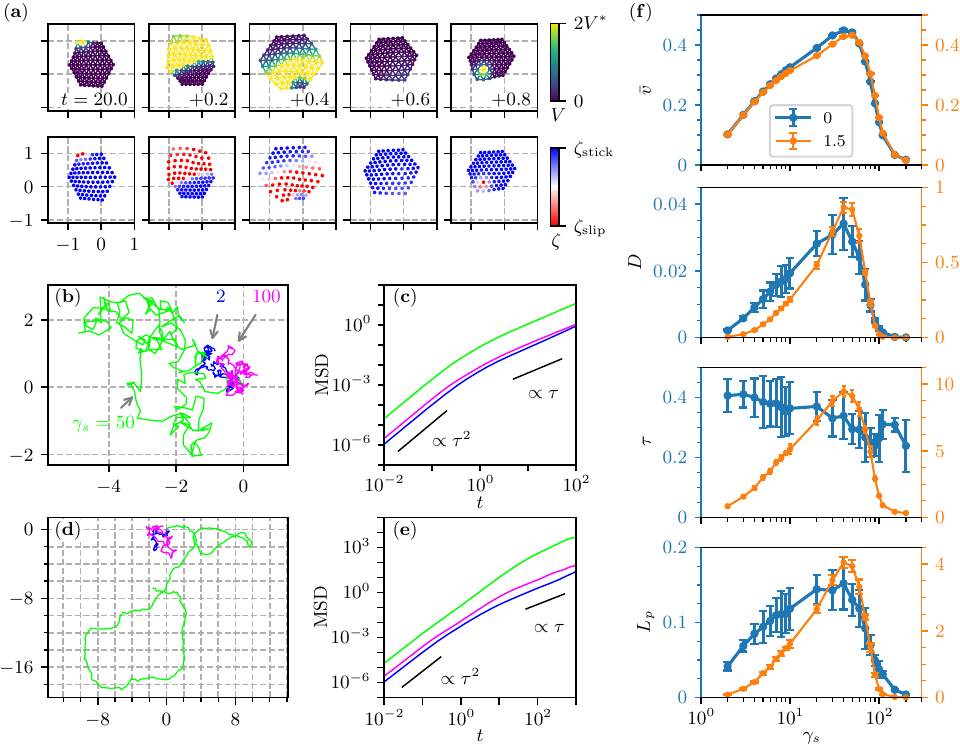}
\caption{Two dimensional crawling motion. (a) Time series of the snapshot of the subcellular elements (top) and adhesion elements (bottom) for $C_m=0$ and $\gamma_s = 50$. The colour of the subcellular elements indicates the value of chemical concentration $V_i$, and that of the adhesion elements represents the value of substrate friction coefficient $\zeta_i$. (b--e) Long-term behaviour of the crawling cell (b,c) without memory $C_m=0$ and (d,e) with memory $C_m=1.5$ on substrates of different stiffness. (b,d) Trajectories of the centre-of-mass position starting from the origin for the same time intervals $\mathit{\Delta}t = 10^{3}$ and (c,e) the corresponding MSD. The colour distinguishes the substrate stiffness $\gamma_s =2$ (blue), 50 (green), and 100 (magenta). (f) Substrate stiffness dependence of crawling characteristics: migration speed $\bar{v}$,  diffusion coefficient $D$, persistence time $\tau$, and persistence length $L_p$. Blue and orange lines distinguish the data without memory $C_m = 0$ (left axis) and with memory $C_m = 1.5$ (right axis).}
\label{fig:fig2}
\end{figure*}
To investigate the impact of the substrate stiffness on the cell migration, first, we analyse the long-time behaviour of the cells without memory effect ($C_m=0$). 
Fig.~\ref{fig:fig2}b shows trajectories of the centre-of-mass position with different substrate stiffness parameters $\gamma_s$ for the same time duration $\mathit{\Delta}t = 10^{3}$. 
The cell migrates randomly due to the perturbations on chemical reactions. 
To quantify the crawling behaviour, we calculate the mean squared displacement (MSD) defined by
\begin{align}
{\rm MSD}(t) = \langle (\bm{X}_{\rm com}(t_{0} + t) - \bm{X}_{\rm com}(t_0))^{2} \rangle 
\label{eq:MSD}
\end{align}
where the average $\langle \cdot \rangle$ is calculated over time $t_0$ and ensembles (10 samples for each parameter set). 

Since the MSD transitions from a short-time ballistic ($\propto t^2$) to a long-time diffusive regimes ($\propto t$) as in Fig.~\ref{fig:fig2}c,
we measure the characteristic speed $\bar{v}$ and diffusion constant $D$ by fitting it using the asymptotic functions: ${\rm MSD}(t) = \bar{v}^2t^2$ for the time interval $t \in [10^{-2},10^{-1}]$ and ${\rm MSD}(t)= 2Dt$ for $t \in [10,10^{2}]$.
From $\bar{v}$ and $D$, we estimate the persistence time $\tau$ and length $L_p$ through the relations $\tau = 2D/\bar{v}^2$ and $L_p = \bar{v}\tau$, respectively.

The results are plotted in Fig.~\ref{fig:fig2}f, where the error bars represent the standard deviation over 10 samples. 
Errors of $\tau$ and $L_p$ are calculated from those of $\bar{v}$ and $D$ using the standard formula of error propagation~\cite{taylor2022}.
Interestingly, the migration speed $\bar{v}$ changes non-monotonically with the substrate stiffness $\gamma_s$. 
That is, it increases first with $\gamma_s$, whereas it decreases for a large substrate stiffness, indicating the existence of optimal substrate stiffness for migration around $\gamma_s = \gamma_s^\ast \approx 50$.
This is consistent with the experimental reports~\cite{Palecek1997,Schreiber2021On,Pallares2023}.
In contrast, the persistence time $\tau$ slightly decreases as $\gamma_s$ increases (Fig.~\ref{fig:fig2}f), which does not agree with the experiments~\cite{Missirlis2014Combined,Novikova2017Persistence-Driven}. 

When the memory effect is switched on ($C_m = 1.5$), the cell migrates more persistently (Figs.~\ref{fig:fig2}d--f). 
In fact, while the characteristic speed is unaltered, the diffusion coefficient and the persistence time and length increase significantly. 
Note that the degradation time of the memory is set to $\tau_m = 1$, which is comparable to the typical period of intracellular chemical travelling wave (see Fig.~\ref{fig:fig2}a). 
Most interestingly, the persistence time is qualitatively different from the case without memory effect and now it also exhibits a non-monotonic change with $\gamma_s$. 
That is, the persistence time increases with the substrate stiffness for $\gamma_s \lesssim \gamma_s^\ast$, which is consistent with the observation~\cite{house2009,oakes2009,Missirlis2014Combined,Novikova2017Persistence-Driven,Ji2023Durotaxis}, whereas it decreases beyond the optimal substrate stiffness $\gamma_s \gtrsim \gamma_s^\ast$. 

\subsection{Intuitive mechanism for non-monotonic speed}\label{subsec:result2}
\begin{figure*}[t]
\centering
\includegraphics[width=0.9\textwidth]{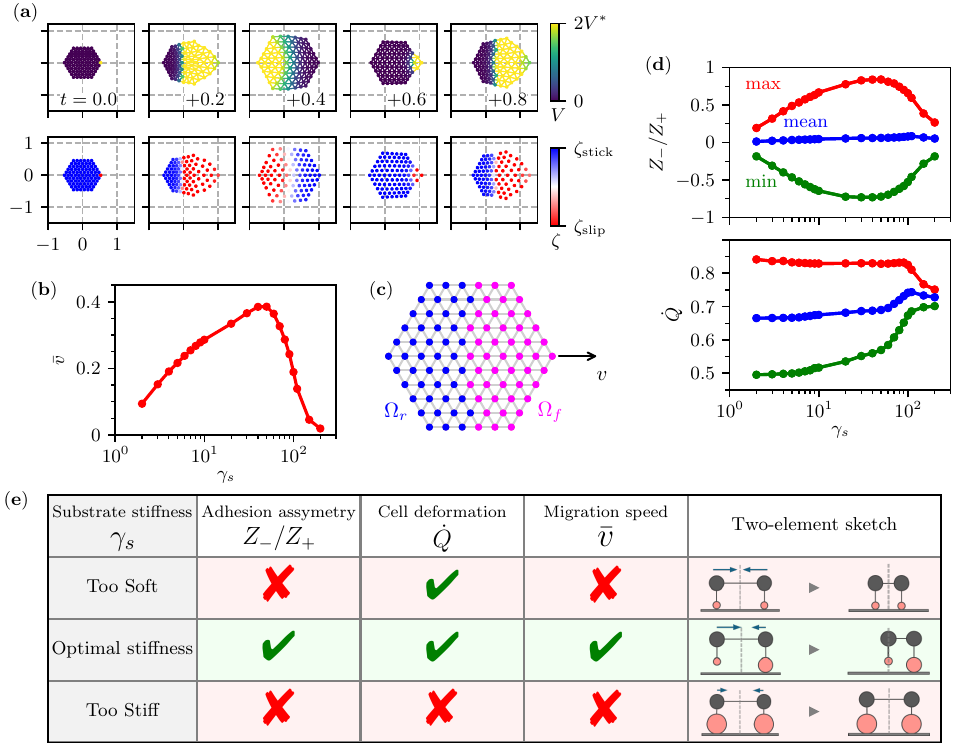}
\caption{One dimensional crawling motion. (a) Time series of the snapshots for $\gamma_{s}=50$. (b) Substrate stiffness dependence of the characteristic migration speed. (c) Two-element approximation, in which the subcellular elements are separated to the front (magenta) and rear (blue) parts with respect to the migration direction. (d) Substrate stiffness dependence of the front-rear asymmetry of substrate friction $Z_-/Z_+$ and the shape deformation $\dot{Q}$. (e) Summary of the intuitive mechanism for the non-monotonic behaviour of characteristic speed on substrate stiffness. }
\label{fig:fig3}
\end{figure*}
Now we ask ourselves from where the optimal substrate stiffness originates. 
To answer this question, we investigate cellular scale mechanism underlying the non-monotonic dependence of characteristic speed on the substrate stiffness. 
Hereafter, we confine ourselves to one-dimensional motion, for the sake of simplicity, by restricting the chemical perturbation to the rightmost subcellular element, which corresponds to the limit of infinitely large memory effect.
This leads to travelling waves propagating leftwards, whereas the cell moves rightwards, i.e., towards $+x$ direction. 
Note that the chemical travelling wave and cell shape deformation are symmetric with respect to $x$-axis (see Fig.~\ref{fig:fig3}a). 
As shown in Fig.~\ref{fig:fig3}b the characteristic speed changes non-monotonically, quantitatively comparable to the migration in two-dimensions.
Therefore, we proceed the analysis based on this one-dimensional migration hereafter.

First, we try to capture intuitive idea on the cellular-scale mechanism. 
To this end, we recall the simplest two-element mechanical model studied in Ref.~\cite{Tarama2018}: 
\begin{align}
\left( \zeta^\prime +\zeta_1(t) \right) v_1 +\xi \ell_0 ( v_1 -v_2 ) &= -\frac{\kappa}{\ell_0} \left( Q - \ell_0 -\ell^{\rm act}(t) \right)\notag\\
\left( \zeta^\prime +\zeta_2(t) \right) v_2 -\xi \ell_0 ( v_1 -v_2 ) &= \frac{\kappa}{\ell_0} \left( Q - \ell_0 -\ell^{\rm act}(t) \right)
\label{eq:N2:EOM}
\end{align}
where $Q = r_1(t) -r_2(t)$. 
Note that we assume the element 1 is existing on the right of the element 2. 
This set of equations are obtained from eqs.~\eqref{eq:Cell mechanics1} in the limit of $N = 2$ and $k_a \to \infty$, where the spatial dimension is reduced to one and the term $\bm{f}^{\rm area}$ is omitted. 
From these, the centre-of-mass velocity is readily obtained as
\begin{align}
\mathcal{V} = \frac{v_1 +v_2}{2}= -\frac{Z_-}{2Z_+} \dot{Q}
\label{eq:N2:V}
\end{align}
with the total dissipation rate $Z_+(t) =  \zeta_1(t) +\zeta_2(t) +2 \zeta^\prime$, the substrate friction asymmetry $Z_-(t) = \zeta_1(t) -\zeta_2(t)$, and the cell deformation 
\begin{align}
\dot{Q} = \frac{-4 Z_+}{ Z_+^2 - Z_-^2  +4 \xi \ell_0 Z_+} \frac{\kappa}{\ell_0} ( Q -\ell_0 -\ell^{\rm act}(t) )
\label{eq:N2:dr12}
\end{align}
This means that the cell velocity is calculated as the product of the substrate friction asymmetry and the force dipole that the cell produces in itself. 
In other words, the cell velocity is approximately given by the front-rear asymmetry in substrate friction ($Z_-/Z_+$) and the magnitude of cell deformation ($\max{\left(Q\right)} -\min{\left(Q\right)}$). 

Based on this simple viewpoint, we analyse how the substrate friction and the cell shape deformation depend on the substrate stiffness by means of two-element approximation. 
To this end, we separate the subcellular elements in our mechanochemical model, eq.~\eqref{eq:Cell mechanics1}, into the front and rear parts with respect to the migration direction (Fig.~\ref{fig:fig3}c) and calculate the average position and total substrate friction of the subcellular  elements and the associated adhesion elements of each part: 
\begin{align}
&\bm{x}_f = \frac{1}{2 N_f} \sum_{ i \in \Omega_f} ( \bm{r}_i  +\bm{r}_{a,i} ), \qquad Z_f(t) = \sum_{i \in \Omega_f} \zeta_i(t)\label{eq:N2:xf}\\&
\bm{x}_r = \frac{1}{2 N_r} \sum_{ i \in \Omega_r} ( \bm{r}_i  +\bm{r}_{a,i} ) , \qquad Z_r(t) = \sum_{i \in \Omega_r} \zeta_i(t)\label{eq:N2:xr}
\end{align}
where $\Omega_f$ ($\Omega_r$) and $N_f = \sum_{i \in \Omega_f}$ ($N_r = \sum_{i \in \Omega_r}$) are the set of subcellular elements of the front (rear) part and its total number, respectively. 
From these, we define 
\begin{align}
Q &= | \bm{x}_f -\bm{x}_r |, \qquad Z_{\pm} = Z_f \pm Z_r
\label{eq:N2:Q_Z_pm}
\end{align}
Since both the cell shape deformation and the substrate friction change in time approximately periodically, we plot the average value of $\dot{Q}$ and $Z_-/Z_+$ as well as the average of their maximal and minimal in each period against the substrate friction $\gamma_s$ in Fig.~\ref{fig:fig3}d. 
For a small $\gamma_s$, while the cell exhibits a large deformation, the substrate friction asymmetry takes only small values, which increases with $\gamma_s$. 
For a large $\gamma_s$, on the other hand, the magnitude of shape deformation and substrate friction asymmetry decrease with $\gamma_s$. 
This observation leads to an intuitive mechanism of the non-monotonic dependence of the characteristic speed on the substrate stiffness (Fig.~\ref{fig:fig3}e). 
That is, on a soft substrate, the cell cannot produce enough asymmetry in substrate friction and, thus, cannot migrate well. 
As the substrate stiffness increases, the substrate friction asymmetry and, thus, the migration speed increase. 
However, too high substrate stiffness suppresses both the substrate friction asymmetry and shape deformation, leading to a decrease in migration speed. 
These effects give rise to the optimal substrate friction for cell crawling. 

\subsection{Cellular-scale mechanism for optimal substrate friction}\label{subsec:result3}
\begin{figure*}[t]
\centering
\includegraphics[width=0.9\textwidth]{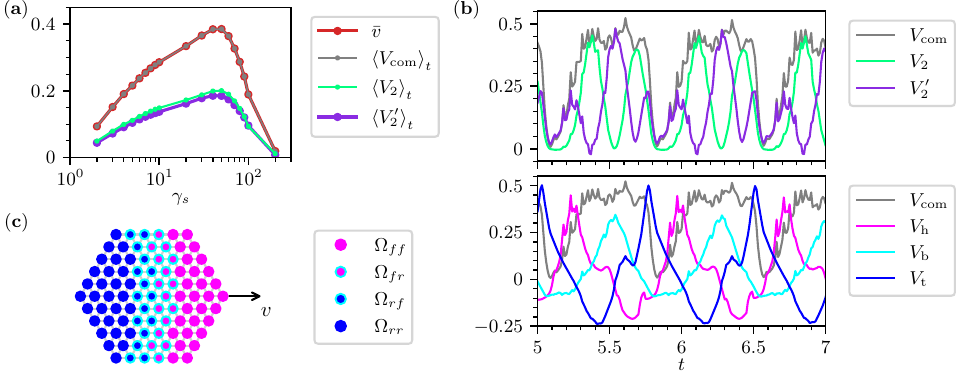}
\caption{(a) Dependence of $\langle V_{2} \rangle_t$ and $\langle V_{2}^\prime \rangle_t$ as well as $\langle V_{\rm com} \rangle_t = \langle V_{2} \rangle_t + \langle V_{2}^\prime \rangle_t$ and $\bar{v}$ on the substrate stiffness $\gamma_s$. (b) Time evolution of $V_{\rm{com}}(t)$, $V_{2}(t)$, $V_{2}^\prime(t)$, $V_f(t)$, $V_c(t)$, and $V_r(t)$. The  substrate stiffness is set to $\gamma_s=50$. (c) Separation of a cell into three regions: head ($\Omega_{ff}$), body ($\Omega_{fr}$ and $\Omega_{rf}$), and tail ($\Omega_{rr}$). }
\label{fig:fig4}
\end{figure*}
So far, we showed the emergence of non-monotonic behaviour in cell crawling against substrate stiffness using a mechanochemical model, and proposed an intuitive mechanism based on a two-element picture. 
Here, we further investigate cellular-scale mechanism theoretically. 
In particular, we analyse the centre-of-mass velocity from eq.~\eqref{eq:Cell mechanics1} based on the two-element approximation. 
For simplicity, we consider the limit $k_a \to \infty$, in which the centre-of-mass velocity consists of two parts: 
\begin{align}
V_{\rm com}(t) = V_2(t) + V_2^\prime(t)
\label{V_com}
\end{align}
See appendix~\ref{sec:Appendix} for the derivation. 
The first term is obtained using $Z_\pm$ and $Q$ defined in eq.~\eqref{eq:N2:Q_Z_pm} as
\begin{align}
V_2(t) = -\frac{Z_-(t)}{2Z_+(t)}\dot{Q}(t)
\label{eq:V_2}
\end{align}
which has the same form as eq.~\eqref{eq:N2:V}. 
The second term represents the deviation from this two-element velocity, and is given by
\begin{align}
V_2^\prime(t) = \frac{Z_r(t) v_r(t) + Z_f(t) v_f(t)}{Z_+(t)}
\label{eq:V_2_prime}
\end{align}
where $v_f$ and $v_r$ are the velocity of front and rear parts, respectively. 

To examine how much contribution these two terms have, we measure $\langle V_2 \rangle_t$ and $\langle V_2^\prime \rangle_t$ with $\langle \cdot \rangle_t$ denoting time average. 
The result is plotted against $\gamma_s$ in Fig.~\ref{fig:fig4}a, which also shows their sum $\langle V_2 \rangle_t +\langle V_2^\prime \rangle_t$ as well as $\bar{v}$. 
Since $\bar{v}$ is the characteristic velocity for a finite $k_a$ plotted in Fig.~\ref{fig:fig3}b, the perfect overlap of $\langle V_2 \rangle_t +\langle V_2^\prime \rangle_t$ with $\bar{v}$ indicates that the characteristic velocity does not depend on $k_a$. 
In addition, the two-element velocity $\langle V_2 \rangle_t$ and the deviation $\langle V_2^\prime \rangle_t$ have the same tendency with $\bar{v}$ with respect to the substrate stiffness $\gamma_s$. 
This indicates that the intuitive explanation based on the two-element picture captures well the basic mechanism behind the non-monotonic behaviour of characteristic velocity and, thus, the origin of the optimal substrate stiffness. 
However, the magnitude of $\langle V_2 \rangle_t$ is only about a half of $\bar{v}$, and its contribution is almost the same as that of deviation $\langle V_2^\prime \rangle_t$. 

Finally, we investigate the physical meaning of the deviation term $\langle V_2^\prime \rangle_t$. 
To this end, we take a closer look at the time evolution of $V_{2}(t)$ and $V_{2}^\prime(t)$. 
Fig.~\ref{fig:fig4}b displays them as well as $V_{\rm{com}}(t)$ for $\gamma_s=50$. 
Within one cycle of $V_{\rm{com}}(t)$, $V_{2}(t)$ hits two peaks, corresponding to the protrusion of the front part (first peak) and the contraction of the rear part (second peak), respectively. 
In contrast, the deviation $V_{2}^\prime(t)$ shows three peaks in one cycle; the first one appears before the motion of the front part (i.e., in front of the first peak of $V_{2}(t)$), the second one between the two peaks of $V_{2}(t)$, and the third one after the motion of the rear part (i.e., behind the second peak of $V_{2}(t)$). 
This suggests that $V_{2}^\prime(t)$ carries the contributions from the motion of three divisions, namely, head, body, and tail of a cell. 
To highlight these contributions, we further separate $V_{2}^\prime(t)$ into three divisions as 
$V_{2}^\prime(t) = V_{h}(t) + V_{b}(t) + V_{t}(t)$, which respectively represent the velocity of cell head ($V_h$), body ($V_b$), and tail ($V_t$). 
See Fig.~\ref{fig:fig4}c, and also appendix~\ref{sec:Appendix} for the derivation and function form of each term. 
As shown in Fig.~\ref{fig:fig4}b, they are basically responsible to the three peaks of $V_2^\prime$, respectively.
This means that the term $\langle V_2^\prime \rangle_t$ includes the contributions from the three divisions, namely the ones from the protrusion of cell head, the flow of body region, and the contraction of tail, which cannot be captured by the two-element approximation. 

\section{Summary and discussions}\label{sec:Discussion}
To summarise, we have demonstrated the cellular-scale mechanism behind adaptive motion of crawling cells in response to substrate stiffness. 
Based on a mechanochemical subcellular-element model, we are successful to reproduce non-monotonic behaviour with respect to the underlying substrate stiffness, indicating the existence of optimal substrate stiffness for cell crawling, which are comparable to the experimental observations. 
In addition, by introducing a memory effect that provides feedback from cell mechanics to intracellular chemical reactions, our model cell exhibits persistence time that also changes non-monotonically and increases with substrate stiffness until the optimal, which agrees with experiments. 
By applying a two-element approximation, we have proposed a mechanism underlying the non-monotonic dependence of characteristic migration speed in terms of the symmetry breaking in cell-substrate adhesion and the cell shape deformation; 
The growth of the front-rear asymmetry in substrate adhesion contributes to the increase in migration speed for a small substrate stiffness. 
In contrast, the shape deformation as well as the adhesion asymmetry are suppressed for a too stiff substrate, leading to the decrease in migration speed. 

In our model, the impact of substrate stiffness is included as the effective adhesion strength based on the fact that the cell-substrate adhesion complex, called focal adhesion, matures on a stiffer substrate~\cite{espina2022}. 
Therefore, the substrate does not deform due to the traction force that are produced by the cells sitting on the substrate. 
There are several studies that consider the deformation of substrate in response to cell traction force~\cite{Arellano2025Multiscale} as well as by external forcing~\cite{Molina2019Modeling}. 

Although our study is based on a simple subcellular-element model, more elaborated continuum models have been proposed to reproduce durotaxis~\cite{Lober2014Modeling}. 
We utilize the particle-based model since it is straightforward to implement the force-free condition.
In addition, our model couples the cell mechanics and intracellular chemical reactions. 
For the chemical reactions, simple Gray-Scott type reaction-diffusion equations are employed so that we can focus on the interplay between the cell mechanics and intracellular chemical reactions.
However, a recent study revealed that the corresponding chemical reaction network is actually more complicated~\cite{Fukushima2019Excitable}. 
Therefore, our model can be modified to include such detailed reaction network, which may enable us to compare more closely with experiments. 

The impact of memory on the cell crawling characteristics depends on both the degradation time of memory and the sensitivity of feedback. 
In this paper, we set the degradation time to unity, which is about the same time as the period of one extension-contraction cycle of the cell observed without the memory effect. 
In contrast, the sensitivity is chosen to make the persistence time for the optimal substrate stiffness comparable to the value (about $10\,{\rm min}$) measured experimentally for Eukaryotic cells~\cite{Li2008}. 
We can analyse more in detail the effect of degradation time and sensitivity theoretically, which we will report elsewhere because it is beyond the scope of this paper.

Finally, in our current study we consider cell crawling on a uniform substrate and do not incorporate spatial variation of substrate stiffness. 
Although our model is applicable to such a situation and actually exhibits cell crawling motion in response to the patterns of substrate, we do not show them here in order to highlight the basic mechanism behind the rigidity sensing of crawling cells. 

In conclusion, we have proposed a mechanism behind the adaptive cell migration in response to the substrate stiffness of a uniform substrate by focusing on cell deformation and symmetry breaking in cell-substrate adhesion.  
We believe that our current theory serves as a prototype to understand more complex durotactic behaviour on patterned substrates with different stiffness.
\begin{acknowledgements}
This work is supported by JSPS KAKENHI Grant Number JP24H01485 and JSPS International Joint Research Program (JRPs) Number JPJSJRP20261608. 
SN thanks to the support by JST SPRING, Japan Grant Number JPMJSP2136.
JSPS Core-to-Core Program ``Advanced core-to-core network for the physics of self-organizing active matter (JPJSCCA20230002)'' is also acknowledged. 
\end{acknowledgements}
\appendix
\section{Two-element approximation}\label{sec:Appendix}
Here we derive the centre-of-mass velocity based on the two-element approximation, eq.~\eqref{V_com} as well as eqs.~\eqref{eq:V_2} and \eqref{eq:V_2_prime}. 
We first eliminate the adhesion elements by substituting eq.~\eqref{eq:Cell mechanics2} to eq.~\eqref{eq:Cell mechanics1} and setting $\bm{v}_i = \bm{v}_{a,i}$ in the limit of $k_a \to \infty$, which yields
\begin{align}
\Xi_{ij} \bm{v}_{j} = \bm{F}_{i}
\label{eq:3.1}
\end{align}
where $\Xi_{ij} = \left(\zeta' + \zeta_{i}(t) + \sum_{k \in \Omega_{i}} \xi \ell\right)\delta_{ij} -\sum_{k \in \Omega_{i}} \xi \ell \delta_{j \ell} \delta_{\ell k}$ and $\bm{F}_i = \sum_{j \in \Omega_i} \frac{\kappa}{\ell_0} \hat{\bm{r}}_{ij} \left\{ r_{ij} - (\ell_0 + \ell_{ij}^{\textrm{act}}(t)) \right\} + \bm{f}_i^{\textrm{area}}$. 
Here $\delta_{ij}$ is the Kronecker delta, that takes 1 if $i = j$ and 0 otherwise. 
Note that $\sum_i \bm{F}_i = 0$ because of the force-free condition. 
Now we project eq.~\eqref{eq:3.1} to the front and rear parts by taking sum over $i \in \tilde{\Omega}_f$ and $i \in \tilde{\Omega}_r$ respectively;
\begin{align}
&\tilde{\Xi}_{ff} N_f \tilde{\bm{v}}_f + \tilde{\Xi}_{fr} N_r \tilde{\bm{v}}_r + \tilde{\bm{w}}_f = \tilde{\bm{F}}_f\label{eq:3.3}\\&
\tilde{\Xi}_{rf} N_f \tilde{\bm{v}}_f + \tilde{\Xi}_{rr} N_r \tilde{\bm{v}}_r + \tilde{\bm{w}}_r = \tilde{\bm{F}}_r
\label{eq:3.4}
\end{align}
where 
\begin{align}
&\tilde{\bm{v}}_\alpha = \frac{1}{N_\alpha} \sum_{i \in \Omega_{\alpha}} \bm{v}_i,~~
\tilde{\bm{F}}_\alpha = \sum_{i \in \Omega_{\alpha}} \bm{F}_i
\label{eq:3.4_vF}\\&
\tilde{\bm{w}}_\alpha = \sum_{j \in \Omega_f}\left( \Xi^{\alpha}_j - \tilde{\Xi}_{\alpha f}\right) \bm{v}_j + \sum_{j \in \Omega_r}\left( \Xi^{\alpha}_j - \tilde{\Xi}_{\alpha r}\right) \bm{v}_j\label{eq:3.4_w}\\&
\Xi^{\alpha}_j = \sum_{i \in \Omega_{\alpha}} \Xi_{ij},~~\tilde{\Xi}_{\alpha \beta} 
= \sum_{i \in \Omega_{\alpha}} \sum_{j \in \Omega_{\beta}} \Xi_{ij}
\label{eq:3.4_Xi}
\end{align}
with $\alpha, \beta \in \{f,r\}$.
$\Omega_f$ and $\Omega_r$ respectively represent the set of subcellular elements in the front and rear parts, and $N_f = \sum_{i \in \Omega_f}$ and $N_r = \sum_{i \in \Omega_r}$ are their total element number.
The set of equations \eqref{eq:3.3} and \eqref{eq:3.4} can be solved for $\tilde{\bm{v}}_f$ and $\tilde{\bm{v}}_r$ as
\begin{align}
&\begin{pmatrix} N_f \tilde{\bm{v}}_f \\ N_r \tilde{\bm{v}}_r \end{pmatrix}=\frac{1}{\mathcal{D}}\begin{pmatrix} \tilde{\Xi}_{rr} & -\tilde{\Xi}_{fr} \\ -\tilde{\Xi}_{rf} & \tilde{\Xi}_{ff} \end{pmatrix}\begin{pmatrix} \tilde{\bm{F}}_f -\tilde{\bm{w}}_f \\ \tilde{\bm{F}}_r -\tilde{\bm{w}}_r \end{pmatrix}
\label{eq:3.4_solution}
\end{align}
where $\mathcal{D} = \tilde{\Xi}_{ff} \tilde{\Xi}_{rr} -\tilde{\Xi}_{fr} \tilde{\Xi}_{rf}$. 
From this, we can compute
\begin{align}
&N_f \tilde{\bm{v}}_f \pm N_r \tilde{\bm{v}}_r =\mathcal{D}^{-1} \left( Z_r \mp Z_f \right) \tilde{\bm{F}}_f \notag\\&-\mathcal{D}^{-1}\left[ \left( \tilde{\Xi}_{rr} \mp \tilde{\Xi}_{rf} \right) \tilde{\bm{w}}_f -\left( \tilde{\Xi}_{fr} \mp \tilde{\Xi}_{ff} \right) \tilde{\bm{w}}_r \right]
\label{eq:3.4_solution_plus_minus}
\end{align}
Here we used the force-free condition and the abbreviations $Z_f = \tilde{\Xi}_{ff} + \tilde{\Xi}_{rf}$ and $Z_r = \tilde{\Xi}_{rr} + \tilde{\Xi}_{fr}$. 
From eq.~\eqref{eq:3.4_solution_plus_minus}, we obtain
\begin{align}
&N_f \tilde{\bm{v}}_f + N_r \tilde{\bm{v}}_r \notag\\&=\frac{ Z_r -Z_f }{ Z_r +Z_f }\left( N_f \tilde{\bm{v}}_f - N_r \tilde{\bm{v}}_r \right)-\frac{2}{ Z_r +Z_f }\left( \tilde{\bm{w}}_f +\tilde{\bm{w}}_r \right)\notag\\&=\frac{ Z_r -Z_f }{ Z_r +Z_f }\left( N_f \tilde{\bm{v}}_f - N_r \tilde{\bm{v}}_r \right)+\frac{2 \left( Z_f N_f \tilde{\bm{v}}_f +Z_r N_r \tilde{\bm{v}}_r \right) }{ Z_r +Z_f }
\label{eq:3.4_solution_plus}
\end{align}
Then, the centre-of-mass velocity is calculated as
\begin{align}
&\bm{V}_{\rm{com}} = \frac{N_f \tilde{\bm{v}}_f + N_r \tilde{\bm{v}}_r}{N}\notag\\
&= \frac{Z_r - Z_f}{Z_r + Z_f} \left( \frac{N_f}{N} \tilde{\bm{v}}_f -\frac{N_r}{N}\tilde{\bm{v}}_r \right) + \frac{ 2 \frac{N_f}{N} Z_r \tilde{\bm{v}}_r +2 \frac{N_r}{N} Z_f \tilde{\bm{v}}_f}{Z_r + Z_f} \notag\\
&= \frac{Z_r - Z_f}{2 \left( Z_r + Z_f \right)} \left( \tilde{\bm{v}}_f -\tilde{\bm{v}}_r \right) 
+ \frac{ Z_r \tilde{\bm{v}}_r + Z_f \tilde{\bm{v}}_f}{Z_r + Z_f} 
\label{eq:3.11}
\end{align}
Here, we set $N_f /N = N_r /N = 1/2$, which is true for an even $N$ and is a very good approximation for a large odd $N$. 
The first term of eq.~\eqref{eq:3.11}, 
\begin{align}
\tilde{\bm{V}}_2 = \frac{Z_r - Z_f}{2 \left( Z_r + Z_f \right)} \left( \tilde{\bm{v}}_f -\tilde{\bm{v}}_r \right) 
\label{eq:3.12_V2}
\end{align}
has the same form as the centre-of-mass velocity of the two-element model in eq.~\eqref{eq:N2:V}. 

To get a physical insight of the second term, we divide the subcellular elements into three divisions, 
namely, head, body, and tail (see Fig.~\ref{fig:fig4}c); 
The head consists of front one third subcellular elements ($\Omega_h$) with respect to the direction of motion, i.e., the front two thirds subcellular elements of the front region ($\Omega_{ff}$). 
The tail consists of the back one third subcellular elements ($\Omega_t$) with respect to the direction of motion, i.e., the back two thirds subcellular elements of the rear region ($\Omega_{rr}$). 
The body consists of the middle one third subcellular elements ($\Omega_b$), i.e., the back one third subcellular elements of the front region ($\Omega_{fr}$) and the front one third of the rear region ($\Omega_{rf}$). 
Then the velocity of the front and rear regions are respectively given by
\begin{align}
\tilde{\bm{v}}_f &= \frac{ N_{ff} \tilde{\bm{v}}_{ff} +N_{fr} \tilde{\bm{v}}_{fr} }{N_f}
= \frac{ N_h \tilde{\bm{v}}_h +N_{fr} \tilde{\bm{v}}_{fr} }{N_f}\label{eq:3.11:v_f}\\
\tilde{\bm{v}}_r &= \frac{ N_{rf} \tilde{\bm{v}}_{rf} +N_{rr} \tilde{\bm{v}}_{rr} }{N_r}= \frac{ N_{rf} \tilde{\bm{v}}_{rf} +N_t \tilde{\bm{v}}_t }{N_r}\label{eq:3.11:v_r}
\end{align}
Here we have defined 
$N_h = N_{ff} = \sum_{i \in \Omega_{ff}}$, $N_{fr} = \sum_{i \in \Omega_{fr}}$, $N_{rf} = \sum_{i \in \Omega_{rf}}$, $N_t = N_{rr} = \sum_{i \in \Omega_{rr}}$, and 
\begin{align}
\tilde{\bm{v}}_{ff} &= N_{ff} \sum_{i \in \Omega_{ff}} \bm{v}_i,~~\tilde{\bm{v}}_{fr} = N_{fr} \sum_{i \in \Omega_{fr}} \bm{v}_i\label{eq:3.11:v_ff}\\
\tilde{\bm{v}}_{rf} &= N_{rf} \sum_{i \in \Omega_{rf}} \bm{v}_i,~~\tilde{\bm{v}}_{rr} = N_{rr} \sum_{i \in \Omega_{rr}} \bm{v}_i
\label{eq:3.11:v_rr}
\end{align}
Therefore, the second term of eq.~\eqref{eq:3.11} is rewritten as
\begin{align}
\tilde{\bm{V}}_2^\prime = \tilde{\bm{V}}_h +\tilde{\bm{V}}_b +\tilde{\bm{V}}_t
\label{eq:3.13_V2prime}
\end{align}
where 
\begin{align}
\tilde{\bm{V}}_h &=\frac{2 Z_f}{3 \left( Z_r +Z_f \right)} \tilde{\bm{v}}_h
\notag\\
\tilde{\bm{V}}_b &=\frac{1}{3 \left( Z_r +Z_f \right)} \left( Z_f \tilde{\bm{v}}_{fr} +Z_r \tilde{\bm{v}}_{rf} \right)
\label{eq:3.13_V3}\\
\tilde{\bm{V}}_t&=\frac{2 Z_r}{3 \left( Z_r +Z_f \right)} \tilde{\bm{v}}_t\notag
\end{align}
Here we set $N_h / N_f  = 2/3$, $N_{fr} / N_f = N_{rf} /N_r = 1/3$, and $N_t /N_r = 2/3$. 
These three velocities represent the velocity of the head, body, and tail divisions, respectively, as shown in Fig.~\ref{fig:fig4}b. 

\end{document}